\documentclass[12pt]{iopart}

\pdfoutput=1 

\usepackage{graphicx}
\usepackage{amssymb}
\usepackage{color}

\usepackage{cite}

\newcommand{\xpct}[1]{\langle{#1}\rangle}    

\newcommand{\dg}{\dagger}
\newcommand{\ve}{\varepsilon}
\newcommand{\beq}{\begin{equation}}        
\newcommand{\eeq}{\end{equation}}
\newcommand{\bea}{\begin{eqnarray}}
\newcommand{\eea}{\end{eqnarray}}
\newcommand{\non}{\nonumber}

\begin{document}

\title{Dynamics of the impurity screening cloud following quantum quenches of the Resonant Level Model }

\author{Shreyoshi Ghosh$^1$, Pedro Ribeiro$^{1,2,3}$, and Masudul Haque$^1$}

\address{$^1$
 Max-Planck-Institut f\"{u}r Physik komplexer Systeme,
  N\"{o}thnitzer Stra{\ss}e 38, D-01187 Dresden, Germany}

\address{$^2$ Russian Quantum Center, 
Business-center "Ural", Novaya street 100A, 
Skolkovo village, Odintsovo district, Moscow area, 143025 Russia}

\address{$^3$ Centro de F\'\i sica das Interac\c c\~oes Fundamentais,
Instituto Superior T\'ecnico, Universidade de Lisboa,
Av. Rovisco Pais, 1049-001 Lisboa, Portugal}

\begin{abstract}
  
  We present a detailed analysis of the evolution of impurity screening cloud in the Resonant Level
  Model (RLM) followed by quenches of system parameters.  The considered quenches are either local
  or found to be effectively local.  The screening cloud is characterized by impurity-bath
  correlators and by the entanglement of a block centered around the impurity with the rest of the
  system.  We consider several local quench protocols, such as, quenches from initially decoupled to
  coupled impurity, quenches between different finite couplings of the impurity, and `detuning
  quenches' involving changing the onsite impurity potential away from or to the chemical potential
  of the bath.  The relevant correlators and the entanglement display versions of the `light-cone'
  effect, since the information about the quench travels through the bath at finite speed.  At long
  times (`inside' the light cone), the impurity screening cloud relaxes exponentially to the final
  equilibrium structure, with the relaxation rate given by the emergent energy scale of impurity
  screening.  Also, snaphots of the time-evolving spatial profile of impurity-bath correlators show
  exponential dependences on space, with the length scale given by the screening length.

\end{abstract}

\maketitle

\section{Introduction} 

The study of real-time evolution of many-body quantum systems driven out of equilibrium is a subject
of intense current interest, partly motivated by remarkable experimental advances in the control of
non-equilibrium conditions, both in cold atomic systems and solid-state setups.  As a result of this
interest, many systems and phenomena in condensed matter physics are being re-examined in
non-equilibrium situations.
Quantum impurity problems have long been a central paradigm in the study of correlated quantum
matter.  Impurity models have contributed to influential physical ideas, such as the Kondo model to
the development of the renormalization group and the Caldeira-Legget model to the study of
decoherence.  Quantum impurity models continue today to be a source of new physical insights, and
appear in a variety of contexts.

It is therefore not surprising that the study of impurity physics out of equilibrium is attracting
an increasing amount of attention.  Much work has by now been done on transport through impurities.  More relevant to
the present work is the interest in real-time dynamics induced by sudden changes (quantum
\emph{quenches}) of impurity Hamiltonians \cite{Nuss2014, Medvedyeva, Lechtenberg2014,
  KleineAnders_PRB2014, Costi_PRB14, RatianiMitra_PRB2010, vonDelft_theory_PRB2012, Schiro_PRB2012,
  Latta, SchiroMitra_PRL2014, HeylVojta_PRL2014, HeylKehrein_JPCM2010, LobaskinKehrein_PRB2005,
  KennesMedenVasseur_PRB2014, VasseurMoore_PRL2014, VasseurHaasSaleur_PRL2013,
  SchillerAnders_PRB2014, SchillerAndrei_PRB2014}.

Generally, an impurity model involves a zero-dimensional object, often a single site or spin,
coupled to a bath.  The spatial structure of the bath is ignored in much of the literature, but this
is exactly the aspect that will concern us in this work: namely the appearance of a spatial
signature of the impurity in the bath.  This is the so-called `impurity screening cloud'.
Single-impurity models often  possess an emergent energy scale.  The most famous is the
celebrated Kondo temperature for the single-impurity Kondo model \cite{kondo, hewson}, but similar
energy scales appear in the single-impurity Anderson model \cite{hewson, Anderson_model_PRB61,
  Anderson_model_scale} and the interacting resonant level model \cite{IRLM_original, IRLM_recent}.
There is a length scale $\xi$ corresponding to this energy scale, which suggests that the bath
surrounding the impurity is affected differently at distances less than $\xi$ from the impurity than
at larger distances $x>\xi$, i.e., there should be a `screening cloud' of radius $\xi$ surrounding
the impurity.
Although the impurity screening cloud is difficult to observe directly experimentally, calculations
have shown that this impurity lengthscale does in fact appear in real-space properties of the bath.
The properties (e.g., persistent current or conductivity) of a mesoscopic device containing a Kondo
or Anderson impurity have been found to behave differently if the device size is larger or smaller
than the size of the Kondo cloud \cite{AffleckSimon_PRL01, SimonAffleck_PRB01, SimonAffleck_PRL02,
  SimonAffleck_PRB03, HandKrohaMonien_PRL06}.  Numerical and variational calculations have found
real-space properties (e.g., impurity-bath correlation functions, distortion of local density of
states, entanglement properties, etc) to be different for $x<\xi$ and $x>\xi$
\cite{BusserAndaDagotto_PRB10, Simonin_arXiv07, Goth_Assaad, Holzner-etal_PRB09, Borda,
  BordaGarstKroha_PRB09, Saleur, MitchellBulla_PRB11, Laflorencie_JSM1, Laflorencie_JSM2, SougatoBose_kondocloud}, for
Anderson and Kondo models and for spin-chain versions of the Kondo model.
These recent results extend earlier perturbative calculations of real-space structure
\cite{BarzykinAffleck_PRB98, ishii, SorensenAffleck96}.

Perhaps the simplest example of an impurity generating a spatially extended screening cloud is that
of the resonant level model (RLM), which is a model of non-interacting spinless fermions.  A single
`impurity' site (or level) is weakly tunnel-coupled to a fermionic bath.  The impurity coupling $J$
generates a small energy scale $\Gamma \propto J^2$, and a corresponding large length scale $\xi
\propto J^{-2}$.  The structure of the equilibrium screening cloud in this model was examined in
detail in Ref.\ \cite{ghosh}.  The screening cloud was characterized primarily through the two-point
correlation function between the impurity site and bath positions at distance $x$ from the impurity.
This impurity-bath correlation was found to have different spatial dependence within ($x<\xi$) and
outside ($x>\xi$) the cloud.  The screening cloud was also described using the entanglement of a
spatial block containing the impurity with the rest of the bath; the dependence of this entanglement
on the distance of the block boundary from the impurity is different when this distance is smaller
than or larger than $\xi$.

In this work, we examine the non-equilibrium response of the screening cloud of the RLM after a
quantum quench.  We start with quenches of the impurity tunnel coupling $J$ (section
\ref{sec_Jquenches}).  This being a \emph{local} quench, the effect of this change propagates
through the bath at a finite speed, and after the `wavefront' has passed through a certain point,
observables related to that position relax to the value corresponding to the ground state of the final
Hamiltonian.  We show that the relaxation of the impurity-bath correlator happens
\emph{exponentially}, with the rate of the exponential decay given by the final impurity screening
energy scale $\Gamma_f \propto J_f^2$.  At any instant, the deviation of the correlators from their
final value is found to increase exponentially with distance from the impurity, with the spatial
length scale $\xi$.  The space- and time-dependence of the correlators happens largely, but not
completely, through the combination ${\Gamma_f}t-x/\xi$.
For quenches starting from the impurity-decoupled case, we can explain many features of the time
evolution of impurity-bath correlators from an analytic calculation in the wide-bandwidth
approximation.  The evolution of the block entanglement also shows numerical signatures of the
screening energy scale $\Gamma_f$, in a somewhat more complicated manner.  Quenches of the impurity
energy (detuning from the bath chemical potential, section \ref{sec_localenergyquench}) results in
similar overall behavior as long as the detuning is moderate.  If the final detuning is nonzero,
this provides an additional oscillation frequency in the dynamics.  Finally, we consider a class of
global quenches of bath parameters (section \ref{sec_global}), and show how these are equivalent to
local quenches for the purposes of screening cloud relaxation.

In the concluding section we provide some discussion and context, and contrast several recent
studies \cite{Nuss2014, Lechtenberg2014, Medvedyeva} which are related in spirit to this work.

\section{Model, tight-binding realization, and equilibrium properties}

We consider  the resonant level model (RLM), given by the Hamiltonian  
\begin{equation} \label{H_RLM} 
H =  \sum_{k} \ve_k c_k^{\dg}c_k   ~+~ \ve_d d^{\dg}d ~-~ J(d^{\dg}c_{x=0} +
c_{x=0}^{\dg}d)  \ .
\end{equation}
Here $c_k$, $c_k^{\dagger}$ ($c_x$, $c_x^{\dagger}$) are the bath fermion operators at momentum $k$
(at position $x$) and $d$, $d^{\dagger}$ are the fermion operators at the impurity site,
$\varepsilon_k$ is the dispersion of the bath fermions, and $J$ is the hopping strength between the
impurity and position $x=0$ of the bath.  The on-site detuning potential $\varepsilon_d$ is
generally tuned to the bath chemical potential, except in Section \ref{sec_localenergyquench}.  Here
$x$ represents the distance from the impurity.  The impurity coupling $J$ is assumed to be much
smaller than the bandwidth of the dispersion $\ve_k$; this is the regime where the physics is
relatively insensitive to details of the bath dispersion.

\begin{figure}
\begin{center}
\includegraphics[width=0.99\linewidth]{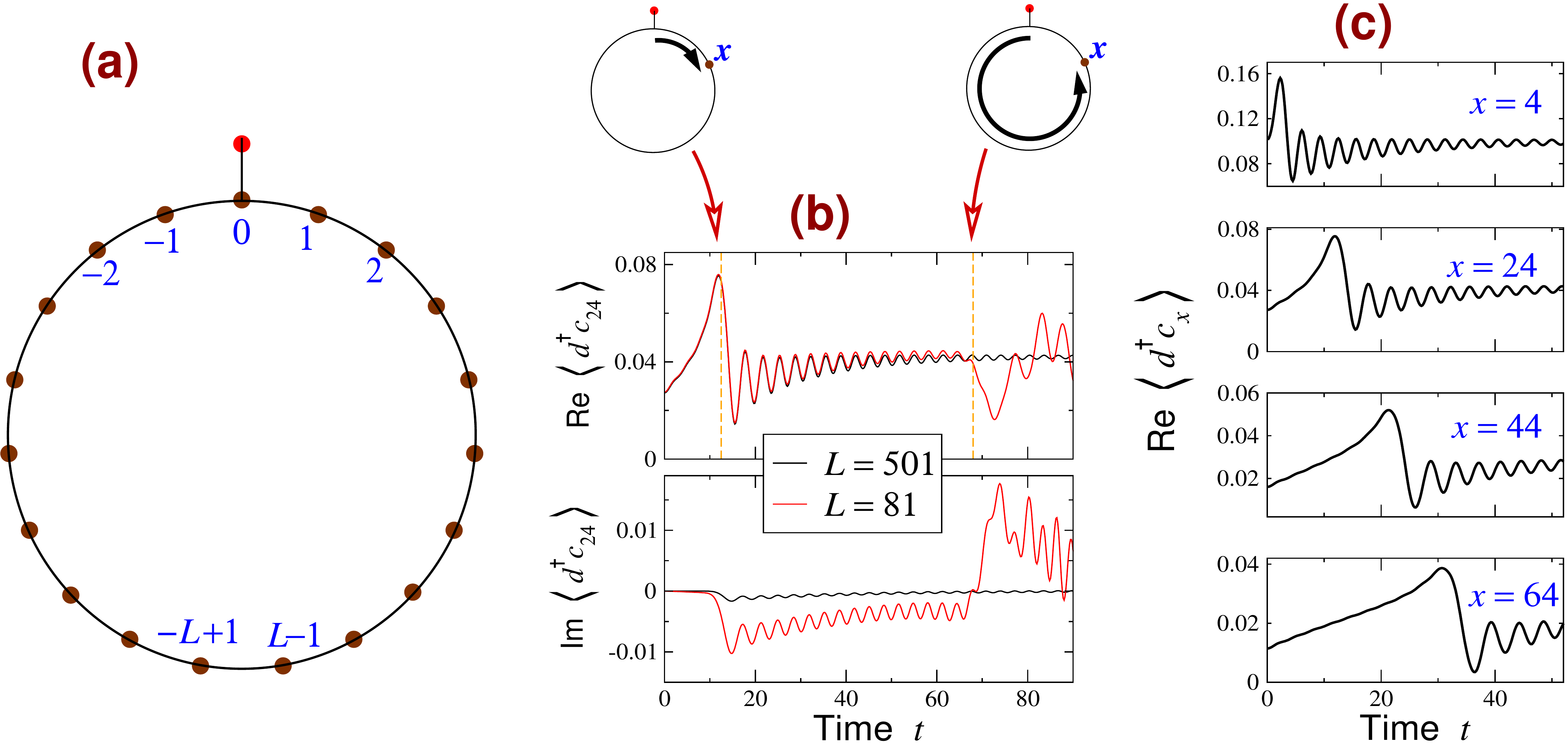}
\end{center}
\caption{ (a) Geometry of tight-binding model (\ref{eq_H_lattice}). (b) Typical behavior of real and
  imaginary parts of $\xpct{d^{\dagger}c_x}$ at $x=24$, after a quench from $J=0.8$ to $J=0.3$. The
  two dashed vertical lines and the associated cartoons indicate the time for the wavefront to reach
  the point x ($\sim{x/2}$), and the time for finite-size effects ($\sim\frac{2L+1-x}{2}$).  (c) The
  propagating wavefront, seen visually by comparing the time evolution of $\xpct{d^{\dagger}c_x}$
  for different $x$. }
\label{introfigs}
\end{figure}

At equilibrium, the RLM for small $J$ displays a screening cloud, as examined in detail in Ref.\
\cite{ghosh}.  The details depend weakly on the nature of the bath; we will concentrate on a
one-dimensional bath.  Specifically, we will use a closed tight-binding chain with $2L-1$ sites,
labeled as $x= -(L-1), \ldots, -1, 0, 1, \ldots, (L-1)$, as shown in Figure \ref{introfigs}(a).  We
will mostly work with nearest-neighbor (n.n.) hopping in the bath, so that the Hamiltonian is
\begin{equation}  \label{eq_H_lattice}
H =  -\sum_{i=-L+1}^{L-1}(c_i^{\dagger}c_{i+1}+ c_{i+1}^{\dagger}c_{i}) 
~+~ \ve_d d^{\dg}d 
~-~ J\left(c_0^{\dagger}d ~+~ d^{\dagger}c_0\right).
\end{equation}
The site label $L$ is identified with site $-L+1$ as usual for periodic boundary conditions. 
We take the bath n.n.\ hopping to be unity, so that energies (time) is measured in units of the bath
n.n.\ hopping strength (inverse of the bath n.n.\ hopping strength).  The dispersion is then
$\epsilon_k=-2\cos{k}$, with $k= 2\pi{n}/(2L-1)$ and $n=-L+1,...,L-1$.  In Section \ref{sec_global},
we also modify the bath by adding next-nearest-neighbor hoppings to the tight-binding chain.  We
will also mostly consider half-filling, i.e., $L$ particles, since the total system including the
impurity has $2L$ sites.  The bandwidth in these units is $4$, we are thus interested in the
parameter regime $J\ll4$.  In practice, we will restrict the final $J$ values to be $\lesssim0.5$.

The equilibrium screening cloud around the impurity is well described by the single particle
impurity-bath correlator $\xpct{d^{\dg} c_x}$ \cite{ghosh}.  The zero-temperature equilibrium
spatial structure of $\xpct{d^{\dg}c_x}$ features a rapid oscillatory behaviour with wavelength
given by the Fermi momentum $k_F$, convoluted with an envelope function that crosses over from a
$\propto {\rm ln}~ x$ behaviour at small $x$, to an algebraic decay $\propto x^{-1}$ at large $x$.
The length scale $x\sim \xi$ at which this crossover happens, i.e., the `size' of the impurity
screening cloud, is $\xi=v_F/\pi J^2 \rho(\ve_F)$, scaling as $J^{-2}$ with the impurity-bath
coupling.  Here $v_F$ the Fermi velocity and $\rho(\ve_F)$ the normalized single-particle density of
states [$\int d\omega \rho(\omega) = 1$] of the bath at the Fermi energy $\ve_F$.  With our units,
at half-filling $v_F= 2$ and $\rho(\ve_F)= 1/(2\pi)$, so that $\xi=4/J^2$.


In Ref.\ \cite{ghosh}, the equilibrium screening cloud was also characterized through a study of the
entanglement entropy, motivated by similar characterizations in Refs.\ \cite{Laflorencie_JSM1,
  Laflorencie_JSM2, SaleurVasseur_PRB13}.  For this purpose, one considers the entanglement entropy
$S_{J}^{[\ell]} = -\mathrm{Tr}_{A_{\ell}} [\rho_{A_{\ell}} \ln \rho_{A_{\ell}}]$ of the subsystem
$A_{\ell}$ of total size $2\ell$ centered around (and containing) the impurity.  Here
$\rho_{A_{\ell}}$ is the reduced density matrix of the subsystem $A_{\ell}$ obtained by tracing out
the degrees of freedom lying outside the subsystem, starting from the wavefunction (in this case the
ground state wavefunction) of the full system.  The so-called impurity entropy, defined as
\begin{equation}
S_{imp}^{[\ell]} = S_{J}^{[\ell]}-S_{J=0}^{[\ell]} ,  \label{S_imp}
\end{equation}
was shown to have, as a function of the subsystem size $\ell$, a crossover  to a an algebraic $\propto \ell^{-1}$ decay at
large $\ell$ around the same characteristic length scale $\xi$ \cite{ghosh}.

In the following we analyse the non-equilibrium dynamics of the screening cloud after a quantum
quench by monitoring the evolution of $\xpct{d^{\dg} c_x}$ and $S^{[\ell]} $ as a function of time.
For a system initialised in the ground-state of the Hamiltonian for $t<0$, we consider local
quenches, where either $J$ or $\ve_d$ are changed abruptly at $t=0$ to their final values, and also
a quench in the global parameters corresponding to an abrupt change of the dispersion relation
$\ve_k$ of the bath.  For a bath with an infinite number of degrees of freedom, local correlators
are expected to relax to the local equilibrium values for asymptotic long times after a local
quench. For these cases we analyse the way the new equilibrium form of the screening cloud is
reached.

\section{Quenches of the impurity coupling $J$ \label{sec_Jquenches}}

In this section we will consider quenches of the impurity-bath coupling $J$.  The system starts in
the ground state for $J=J_i$ and, after the $J_i \to J_f$ quench,  evolves under the final
Hamiltonian with  $J=J_f$.  In subsections \ref{subsec_Jquenches_general},  \ref{subsec_Ji_0},
\ref{subsec_Ji_finite}, we  consider the evolution of the  equal-time impurity-bath correlation function
\begin{equation}
C(x,t) = \left\langle d^{\dagger}(t)c_{x}(t)\right\rangle   . 
\end{equation}
In subsection \ref{subsec_entanglement} we consider the evolution of the block entanglement entropy.

\subsection{Time evolution of impurity-bath correlators: general features \label{subsec_Jquenches_general}}

In figures \ref{introfigs}(b) and \ref{introfigs}(c) we display some general features of the time
evolution of the impurity-bath correlator, highlighting effects of locality, information
propagation, and the appearance of infinite-size quench features in finite-size systems.

After a local quench, information about the change in the system should propagate out from the point
of disturbance with speed $v_F$.  This effect has been discussed widely in recent times in terms of
the so-called Lieb-Robinson bound and the `light-cone' effect, for both local and global quenches
\cite{LiebRobinson_1972, Nachtergaele_JSP2006, CalabreseCardy_PRL2006, Nuss2014, Medvedyeva,
  Lechtenberg2014, LewensteinTagliacozzo_JSM2014, DivakaranIgloiRieger_local_JSM2011,
  StephanDubail_JSM2011, ChenauBlochKollath_Nature2012, EsslerLauchli_PRL2014, SchuchEisert_PRA2011,
NachtergaeleSims_CMP2006, BravyiHastingsVerstraete_PRL2006, Manmana_etal_PRB2009, Cazalilla_PRL2006}. 
For local quenches, one typical way of formulating this causality effect is that a local observable
at position $x$ will be substantially affected by a local quench at position $y$ only in the
space-time region $t>|x-y|/v$ (`inside' the light cone).  Here $v$ is the characteristic speed.

The equal-time impurity-bath correlator $C(x,t)= \xpct{d^{\dagger}c_x}$ is not a local observable.  Thus, it
evolves in time also outside the light cone, $t<x/v_F$, but this part of the dynamics is entirely
due to the $d$ operator and hence relatively simple.  The dynamics is dramatically affected at
$t\sim{x/v_F}$, when the `wavefront' reaches the point $x$.  This is shown in the examples of Figure
\ref{introfigs}(b), in particular the top left schematic.  The panels in Figure \ref{introfigs}(c)
visually display the increasing time required for the wavefront or signal to reach farther points.

Within the light-cone, i.e., for $t>{x/v_F}$, the correlator decays approximately (upto
$\sim{L}^{-1}$ effects) toward its final ground-state value, $C_f^{\rm Eq}(x,t)$.  The
reasoning behind this is that a local quench is a ${L}^{-1}$ effect and therefore does not inject a
thermodynamically significant energy into the system, hence a large enough  system should relax
approximately to its ground state.  Figure \ref{introfigs}(b) also shows that finite-size effects
are strongly felt when the signal or wavefront traverses the other end of the system and then
reaches point $x$ (top right schematic).  Until this point of time, $t \approx(2L-x)/v_F$, the
correlator $\xpct{d^{\dagger}c_x}$ is approximately $L$-independent, thus by analyzing this region
we essentially access the behavior of $\xpct{d^{\dagger}c_x}$ in an infinite system.  By increasing
the ring size $L$, we get larger windows of time where the infinite-system dynamics can be accessed.

The details of the dynamics depend sensitively on the parameters, filling, and the site $x$.  Even
for the half-filling case (where the behavior is cleanest, also in equilibrium), the real and
imaginary parts of $C(x,t)$ show different dynamics for odd and even $x$.  For all $x$, the
imaginary part starts from zero and (for large enough systems) goes to zero at long times. For even
distances $x$, e.g., Figure \ref{introfigs}(b), the imaginary part shows strong finite-size effects
and for large enough systems is almost time-independent.  For odd $x$ (not shown), the behavior is
reversed: $\mathrm{Re}[C(x,t)]$ shows strong $L$-dependence while $\mathrm{Im}[C(x,t)]$ does not.
In order to avoid these system- and $x$-dependent details, we will mostly show absolute values,
$|C(x,t)|$ or $|C(x,t)-C_f^{\rm Eq}(x,t)|$. 

A local quantity such as $\xpct{c^{\dagger}_xc_x}$ or $\xpct{c^{\dagger}_xc_{x+1}}$ shows a clearer
light-cone effect, i.e., stays constant for $t<{x/v_F}$ until the wavefront reaches $x$.  Such data
is not shown in this work, as our focus is the impurity screening cloud, which is better embodied in
the impurity-bath correlator $C(x,t)= \xpct{d^{\dagger}c_x}$.  Also, Ref.\ \cite{Medvedyeva} has
discussed the appearance or absence of the light-cone effect in different \emph{unequal}-time
correlators.


\begin{figure}
\begin{center}
\includegraphics[width=0.7 \linewidth]{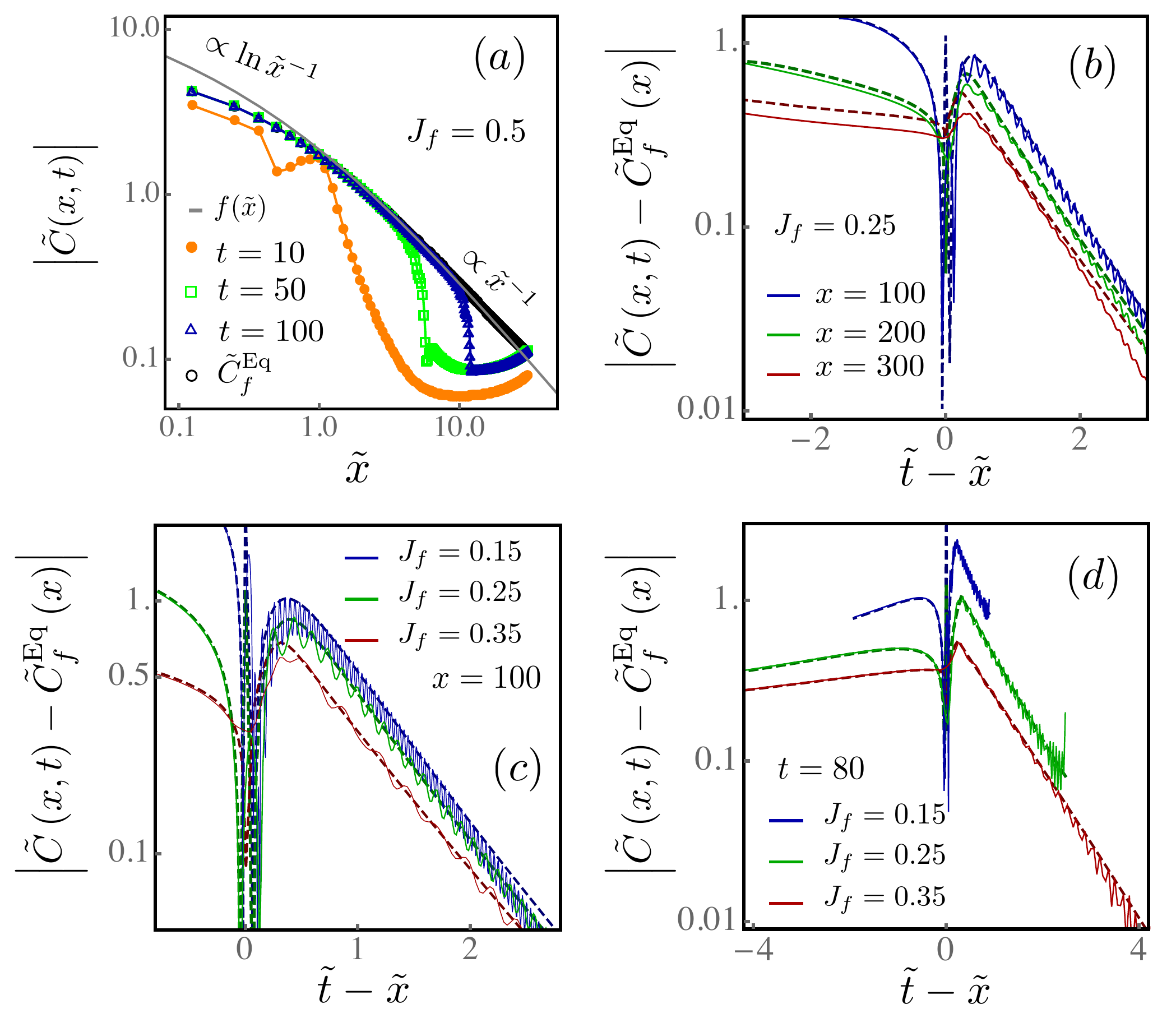}
\end{center}
\caption{ Quenches from an initially decoupled impurity  ($J_i=0$).  
The solid lines are tight-binding numerical results for  sizes $L \simeq 1000$.  Dashed lines are
the infinite-bandwidth infinite-size results of  Eq.\ (\ref{dcanalytic}). 
(a) Snapshots of the rescaled impurity-bath correlation function $\tilde C(x,t)$ as a function of
$\tilde x =x/\xi_f$ for different times.  $f(\tilde x)$ is the scaling function obtained in
Ref. \cite{ghosh}.  (b,c) Time evolution of the difference between $\tilde C(x,t)$ and its
equilibrium value $\tilde C_f^{{Eq}}(x)$, at several positions $x$. We plot against the rescaled and
shifted time variable $\tilde t - \tilde x$.  (d) Snapshots of the spatial profile of $\left| \tilde
  C(x,t)-\tilde C_f^{{Eq}}(x) \right|$ at time $t=100$.    }
\label{fig1}
\end{figure}

\subsection{Starting from decoupled impurity: $J_{i}=0$  \label{subsec_Ji_0}}

We first consider an initially decoupled impurity level which is suddenly coupled at time $t=0$ by
changing the impurity hopping parameter $J$ from $J_i=0$ to finite $J_f$.  The initial correlation
functions are thus given by
\begin{equation}
\xpct{c_k^{\dg}c_{k'}}_0=\delta_{kk'} \Theta(k_F - |k| );  \quad \xpct{d^{\dg}d}_0=n_d;  \quad
\xpct{d^{\dg}c_x} = 0 \label{eq_J0_initcorr} 
\end{equation}
where $n_d$ is the initial occupancy of the impurity level. 
%
%
In the final equilibrium state, with $J=J_f \ne 0$, the impurity is screened over a finite distance
$\xi_f = v_F/ \Gamma_f$, with $\Gamma_f= \pi J^2_f \rho(\ve_F)$.

In these conditions the evolution of the equal-time two-point correlator $\xpct{d^{\dg}(t)c_i(t)}$
is amenable to an analytic treatment along the same lines of Ref.\ \cite{Medvedyeva}.  The approach
is detailed in the Appendix.  We will express results in terms of the rescaled correlator
\begin{equation}
C(x,t) = \frac{J}{\pi^{2}v_{F}}\tilde{C}(x,t)  \, .
\end{equation}
In the thermodynamic limit and the large-bandwidth limit, we obtain the following expression for the
time evolution of the rescaled impurity-bath correlator:
\begin{eqnarray}
\fl \tilde{C}(x,t)  =     \pi  \int_{-\infty}^0 d \varepsilon \frac{ \left[1-e^{-i (\varepsilon -i) \tilde{t}}\right] \cos\left[\tilde{x} \left(\varepsilon +\tilde{k}_F\right)\right]
}{ (\varepsilon -i)} \nonumber \\
   ~-~ \theta(\tilde{t}-\tilde{x})    \left[i\pi  e^{i k_F x} \right]
\int_{-\infty}^0 \frac{d\varepsilon e^{i\varepsilon\tilde{x}}}{\varepsilon^2+1}  \left[ 1-e^{-i
      (\varepsilon -i) \tilde{t}}\right]  \left[  1- e^{i (\varepsilon +i) (\tilde{t}-\tilde{x})}
  \right]       \, . 
 \label{dcanalytic}
\end{eqnarray}
Here $\tilde{x}= x/\xi_f$, $\tilde{t}=\Gamma_f{t}$, and $\tilde{k}_F=\xi_fk_F$ are dimensionless
versions of distance, time and Fermi momentum, obtained by scaling with the length scale $\xi_f$ and
energy scale $\Gamma_f$ of the impurity screening cloud corresponding to the final impurity coupling
$J_f$.  For our lattice Hamiltonian (\ref{eq_H_lattice}) at half-filling, $k_F=\pi/2$.

We comment on several features in this expression:
\begin{itemize}
\item There is a clear and natural separation between behaviors for $\tilde{t}< \tilde{x}$ and for
  $\tilde{t}> \tilde{x}$, i.e., inside and outside the light cone.  The first term shows that the
  evolution is not frozen for $\tilde{t}< \tilde{x}$ (outside the light-cone). As discussed in the
  previous section, this is expected in the impurity-bath correlator because of the evolution of the
  impurity operator $d^{\dagger}(t)$
\item The impurity screening cloud (associated with the ground state of the final Hamiltonian) sets
  time- and length- scales for some of the dynamics, since time and position appear in the
  combinations $\tilde{t}=\Gamma_f{t}$ and $\tilde{x}=x/\xi_f$ in the expression.  In particular,
  the non-oscillatory part of the time-dependence factorizes out of the integrals and is of the form
  $e^{-\tilde{t}}$, which shows that the long-time decay to the final value is exponential in time
  with time constant $1/\Gamma_f$.
\item The time-dependent correlator displays $k_F$ oscillations in space, due to factors like
  $e^{ik_Fx}=e^{i\tilde{k}_F\tilde{x}}$, just like the equilibrium behavior of the correlator
  \cite{ghosh}.
\item In the $\tilde{t}> \tilde{x}$ part (inside the light-cone), a non-oscillatory exponential
  spatial dependence $~e^{\tilde{x}}$ can be factored out of the integral.  Thus, the deviation from
  the final equilibrium value increases exponentially with distance, in the spatial region in which
  the wavefront has already propagated through.
\item 
The long-time limit ($t\to\infty$) of this expression reduces to the equilibrium expression
  for the impurity-bath correlator derived in Ref.\ \cite{ghosh}, showing that for large sizes the
  impurity-bath correlator reduces to the final equilibrium profile. 
\item There are no temporal oscillations with frequency $=2$ (twice the bath hopping or half the
  bath bandwidth).  In calculations with the tight-binding model \ref{eq_H_lattice}, oscillations at this
  frequency appear, as in Figure \ref{introfigs}(b,c); this represents the energy width of occupied
  single-particle states.  However, since the expression is obtained in the infinite-bandwidth
  limit, the bandwidth or hopping parameter of the bath do not appear as an oscillation time-scale.
\end{itemize}

Figure \ref{fig1} shows the impurity-bath correlation function $\tilde{C}(x,t)$ after quenches
starting from the decoupled case, $J_i=0$, comparing numerical tight-binding results using
Hamiltonian (\ref{eq_H_lattice}) against predictions from Eq.\ (\ref{dcanalytic}) above derived in
the large-size large-bandwidth limit.

Figure \ref{fig1}(a) shows snapshots of the spatial profile of $|\tilde{C}(x,t)|$, calculated for the
tight-binding Hamiltonian at different time instants $t$ after the quench.  The spatial profile is
seen to approach the final equilibrium curve at large times.  The (final) equilibrium profile
approximates the analytic equilibrium curve $f(\tilde{x})$, but does not match it exactly for a
single finite system size and single $J_f$ value, as discussed in detail in Ref.\ \cite{ghosh}.

To highlight the approach to equilibrium, in Figures \ref{fig1}(b,c,d) we show the correlator with
the final equilibrium value subtracted off, i.e, $\tilde{C}(x,t)-\tilde{C}_f^{{Eq}}(x)$.  Motivated
by the light-cone structure, we plot this as a function of the variable $\tilde t - \tilde x$.  The
behavior outside the light-cone ($\tilde{t}<\tilde{x}$) and inside the light-cone
($\tilde{t}>\tilde{x}$) look sharply different in this quantity.  The quantity is plotted for
different bath positions $x$ (Figure \ref{fig1}(b)), and for different final coupling $J_f$ at the
same bath position $x=100$ (Figure \ref{fig1}(c)).  In the log-linear plots, the decay for
$\tilde{t}>\tilde{x}$ appear as straight lines with the same slope in each of these cases, showing
that the decay to the final value inside the light cone happens exponentially with the time scale
associated with the (final) impurity screening cloud.

Figure \ref{fig1}(d) plots spatial snapshots of the correlator at the same time instant, $t=100$.
We choose to plot it against the same variable $\tilde{t}-\tilde{x}$ used for showing the time
evolution in Figures \ref{fig1}(b,c).  The similarity of panel \ref{fig1}(d) with the plots in
\ref{fig1}(b,c) highlights the fact that the dynamical features are dominanted by the quantity
$\tilde{t}-\tilde{x}$.  In Figure \ref{fig1}(d) we clearly see an exponential \emph{increase} of the
deviation from the final value, with increasing distance.  This is consistent with the
$e^{\tilde{x}}$ factor in the infinite-bandwidth expression, discussed above.  There are small
$e^{ik_Fx}$ oscillations, but they are negligible everywhere except at small values of $x$.  These
oscillations appear in both the tight-binding data and in the infinite-bandwidth equation, and show
strong finite-size effects.  

In each  panel of Figure \ref{fig1}, we show tight-binding results (solid curves) together with
the prediction Eq.\ \ref{dcanalytic} (dashed curves).  For the regime under discussion, the
infinite-bandwidth continuum approximation captures well the essential features of the impurity
cloud dynamics as seen through the $C(x,t)$ correlator.  The agreement is generally excellent.
Points of disagreement are the temporal oscillations of frequency 2 (half the bandwidth) appearing
in the lattice data but not in the infinite-bandwidth analytics, and the difference between the
$e^{ik_Fx}$ oscillations, which are anyway very small.


\subsection{Quench from finite $J_i$ to finite $J_f$ \label{subsec_Ji_finite}}

\begin{figure}
\begin{center}
\includegraphics[width=\linewidth]{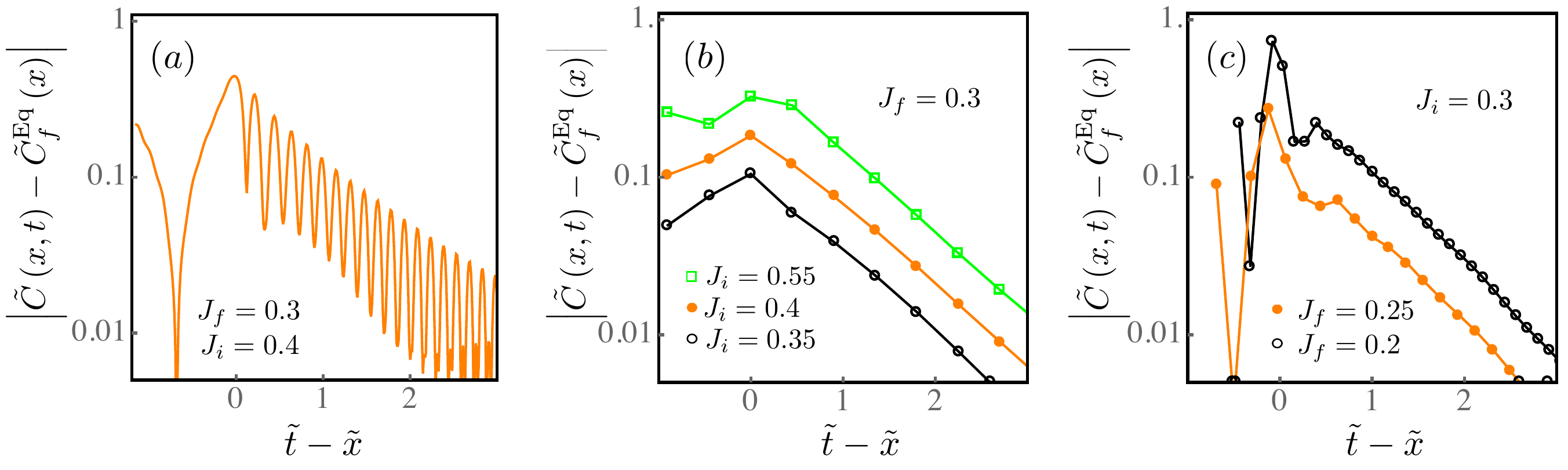}
\end{center}
\caption{Quenches from finite $J_i$ to finite $J_f$. 
(a) Relaxation of the  impurity-bath correlation function, for $x=50$.
In (b,c), the results are averaged over small time intervals $\Delta{t}$ to average out the strong
oscillations  and emphasize the exponential decay with  rate $\Gamma_f$.  In panels (b) and (c), $\Delta{t}=20$ and
$\Delta{t}=12$  respectively. 
}
\label{fig2}
\end{figure}

We now consider quenches from finite $J_i\ne 0$ to a different finite value $J_f$.  The initial
correlators now no longer have a simple form like Eq.\ (\ref{eq_J0_initcorr}) for the $J_i=0$ case,
but rather they describe a screening cloud with characteristic length  $\xi_i\propto{J_i}^{-2}$.

After the quench, the correlator $C(x,t)$ relaxes in an analogous way to the new equilibrium profile
corresponding to the length $\xi_f\propto{J_f}^{-2}$.  As the light cone spreads, $C(x,t)$ within
the light cone ($x<v_Ft$) at time $t$ will relax exponentially to the final profile.

In principle, an integral expression for the correlation function $C(x,t)$ could be derived
analytically using a similar method to the one given in the appendix for $J_i=0$.  However, such an
expression would involve additional integrals, because the initial $C(x,0)$ is already nontrivial
\cite{ghosh}.  We therefore restrict to showing numerical results from the tight-binding model.  For
$J_i\ne 0$, the temporal oscillations of $C(x,t)$ (with frequency $\ve_F$) are much stronger than in
the $J_i=0$ case, as seen in Figure \ref{fig2}(a).  To focus on the exponential relaxation, in
Figures \ref{fig2}(b) and \ref{fig2}(c) we average out the oscillations by plotting the running
average of $C(x,t)$, i.e., each point is an average of $C(x,t)$ taken over a small time window
$\Delta t$ around time $t$.  For a variety of $J_i$ and $J_f$ values, the relaxation is seen to be
always exponential, with the time scale corresponding to the final impurity screening scale
$\Gamma_f= \pi J^2_f \rho(\ve_F)$.

\subsection{Entanglement entropy  \label{subsec_entanglement}}

\begin{figure*}
\begin{center}
\includegraphics[width=\textwidth]{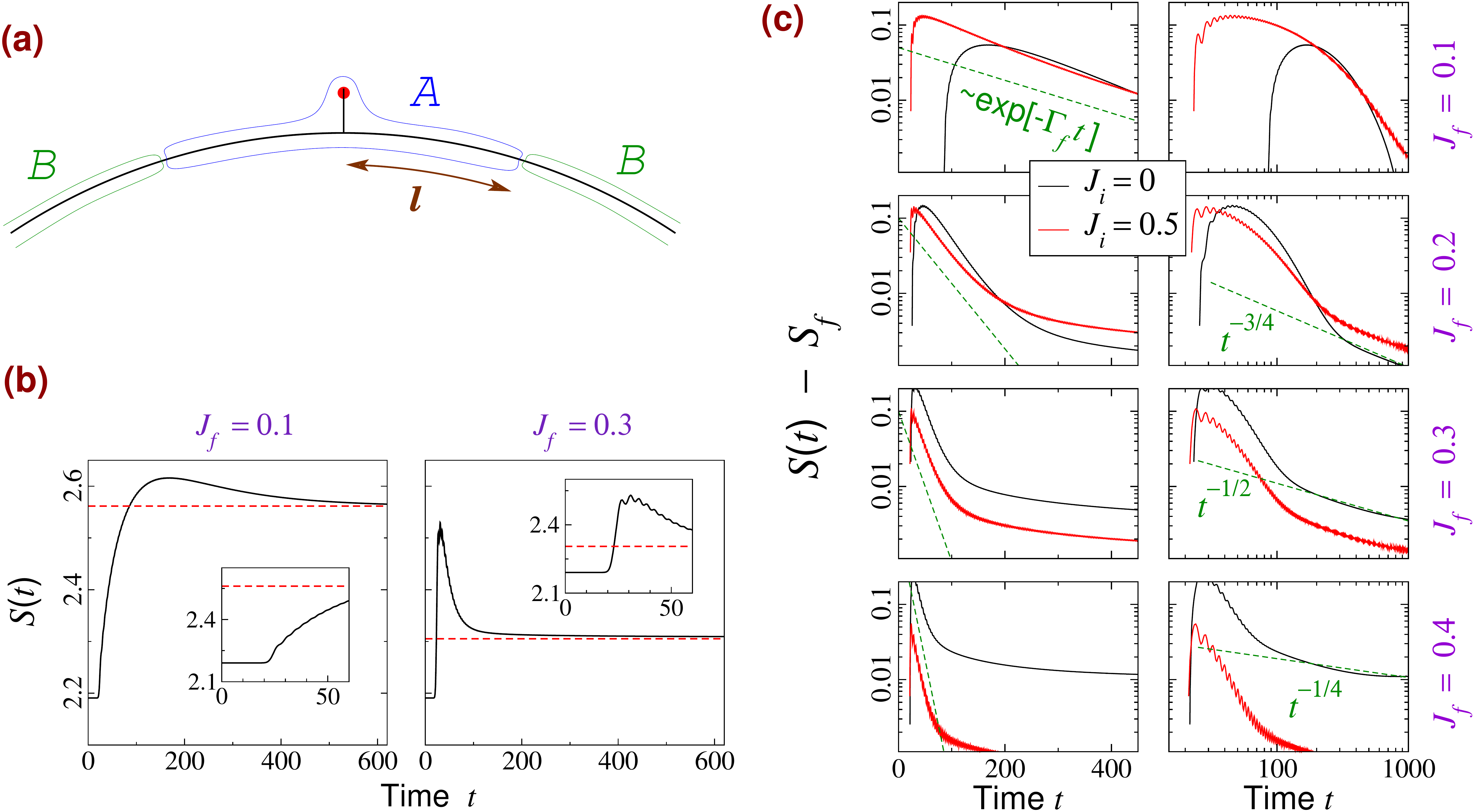}
\end{center}
\caption{(a) Geometry for defining block entanglement between $A$ and $B$ partitions.  (b) Time
  evolution of entanglement entropy, with $l=40$, after a quench from $J_i=0$ to two different final
  $J_f$.  The dynamics starts at $t=l/v_F= 20$, after the signal reaches bourndary between $A$ and
  $B$ blocks.  The insets zoom around this instant.  The dynamics clearly is faster for larger
  $J_f$.  The entanglement entropy approaches the final ground-state value (dashed horizontal lines)
  at long times.  (c) Effect of $J_f$ on the relaxation explored in detail, for quenches starting
  from $J_i=0$ and from $J_i=0.5$.  The left panels (log-linear scale) highlight the initial
  ${\sim}e^{-\Gamma_ft}$ behavior.  The right panels (same data, log-log scale) show that the
  long-time behavior is an approximate power law, with drifting exponent that suggests a logarithmic
  correction as in Ref.\ \cite{Eisler_Peschel}.  }
\label{fig_entng}
\end{figure*}

We now consider the time evolution of the entanglement entropy $S^{[\ell]}(t)$ between a block $A$
containing the impurity at the center, and the rest of the system, which we call $B$.  The geometry
is shown in Figure \ref{fig_entng}(a).  The block $A$ contains a total of  $2\ell$ sites including
the impurity, so that the boundary between the blocks is at distance $\ell$ from the impurity. 

The entanglement entropy in a free-fermion system can be obtained from the matrix of all two-point
correlators within one of the blocks \cite{Peschel_JPA2003, Eisler_Peschel, ghosh}: if $\nu_i$'s are
the eigenvalues of this matrix, then $S=-\sum_i [\nu_i{\rm ln} \nu_i+(1-\nu_i) {\rm ln} (1-\nu_i)]$.
In the present case, we calculate the entanglement $S^{[\ell]}(t)$ at instant $t$ from the matrix of
equal-time two-point correlators at time $t$. 

Ref.\ \cite{Eisler_Peschel} has studied the entanglement evolution after a particular local quench,
for a block containing the quench position at the center.  (The strength of a single bond of the
chain is quenched.)  The results are intriguing and, to the best of our knowledge, not yet
completely understood: the block entanglement entropy decays to the final equilibrium value with a
power law modified logarithmically.  This is accompanied by $\ln{t}$ behaviors of some of the
eigenvalues $\nu_i$ of the correlation matrix.  Our situation is broadly similar, but we are
particularly interested in the role of the impurity screening scale in the dynamics.

Figure \ref{fig_entng}(b) shows the behavior of $S^{[\ell]}(t)$ after a quench starting at $J_i=0$.
(The overall behavior for other $J_i$ values is very similar.)  The entanglement between the blocks
stays frozen at the initial value until the information wavefront reaches the boundary between the
blocks, i.e., until $t\sim \ell/v_F$.  This may seem surprising because many correlators in the
correlation matrix of $A$ have dynamics at smaller times (as seen in previous sections), but it is
physically reasonable because $S^{[\ell]}(t)$ measures the entanglement between $A$ and $B$ blocks,
and the $B$ block cannot know about the local quench until this time.  After a sharp reaction
shortly after $t\gtrsim \ell/v_F$, the entanglement relaxes toward the final ground state value,
shown in Figure \ref{fig_entng}(b) with dashed horizontal lines.  The overall form of
$S^{[\ell]}(t)$ is broadly similar to that seen in the local quench of Ref.\ \cite{Eisler_Peschel}.
The response and relaxation behavior are clearly faster at larger $J_f$, consistent with the fact
that the energy scale of the screening cloud ($\Gamma_f$) increases with $J_f$. 

The relaxation behavior is examined in more detail in Figure \ref{fig_entng}(c).  There are two time
regimes, separated by a time scale proportional to $\Gamma_f\sim{J_f}^2$.  (For $J_f=0.1$, only the
first regime is visible.)  The initial relaxation regime is exponential, with the time constant
given by $\Gamma_f^{-1}$.  This feature is related to the physics of impurity screening and does not
appear in the quench of Ref.\ \cite{Eisler_Peschel}.  At later times, the relaxation is slower; the
data for accessible sizes is consistent with a power law but the best-fit exponents vary.  This
suggests that the exact behavior in this long-time regime may well have logarithmic corrections as
in Ref.\ \cite{Eisler_Peschel}.  The time evolution of the eigenvalues $\nu_i$ also shows
logarithmic evolution of some eigenvalues as in Ref.\ \cite{Eisler_Peschel}; a full understanding is
lacking at present, but presumably the $\ln{t}$ features are not related to impurity screening
physics.

\section{Quenches of the impurity detuning $\varepsilon_d$  \label{sec_localenergyquench}}

\begin{figure}
\begin{center}
\includegraphics[width=0.99\linewidth]{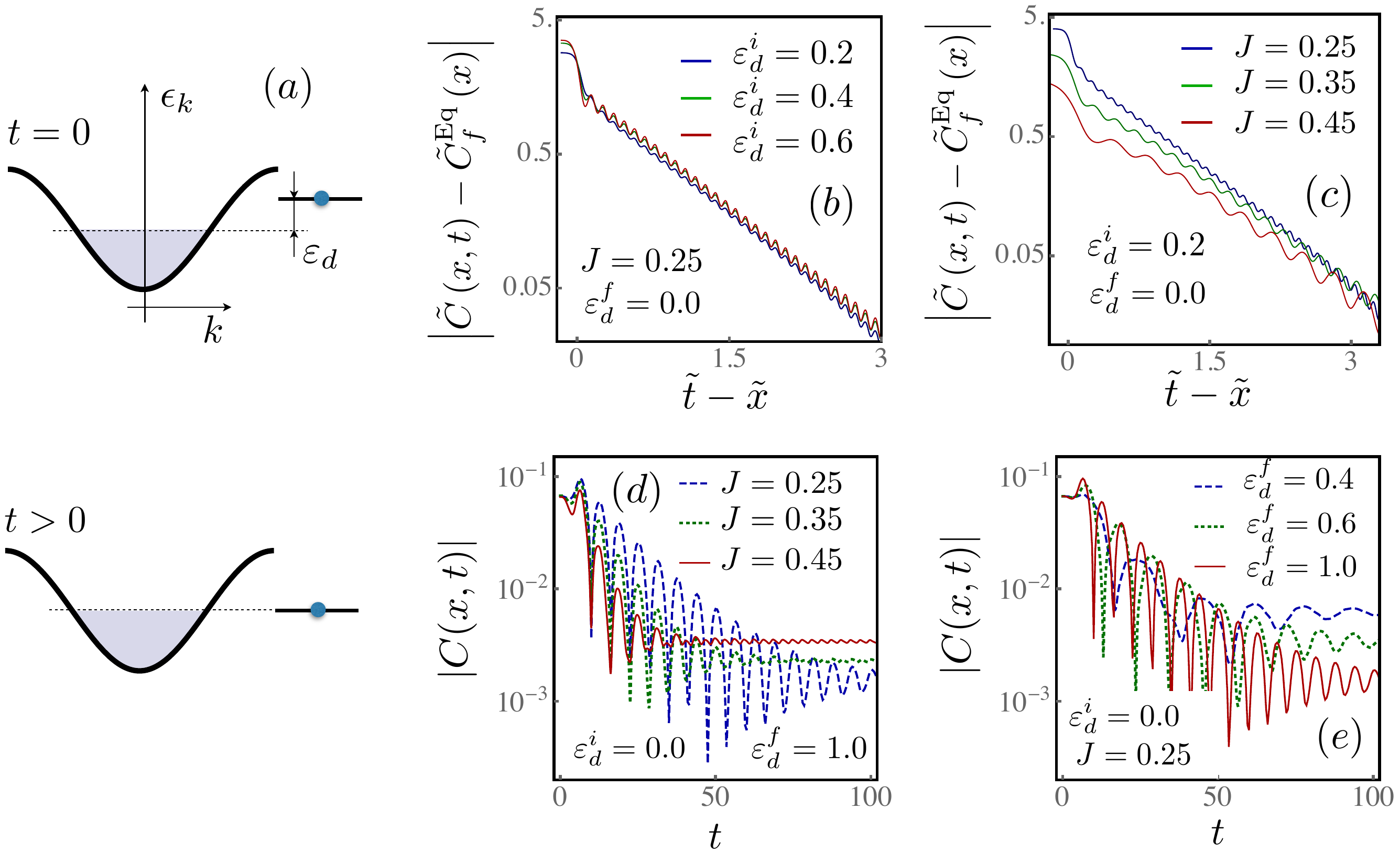}
\end{center}
\caption{ 
(a) Schematic showing a quench from finite to zero detuning. 
(b,c) Relaxation of impurity-bath correlators for quenches to zero detuning. 
(d,e) Evolution of impurity-bath correlators for quenches to finite detuning.  In order to
compare frequencies, in these panels we plot against time without rescaling.  
}
\label{fig_detuning}
\end{figure}

In this section we study the dynamics after quenching the detuning, i.e., the onsite energy bias of
the impurity site, denoted by $\varepsilon_d$ in Equations (\ref{H_RLM}) and (\ref{eq_H_lattice}).  The
detuning energy is measured with respect to the Fermi energy, as shown in Figure
\ref{fig_detuning}(a).  We consider both quenches from an initially detuned impurity potential to
the zero detuning case, $\ve_d^i\ne0$ to $\ve_d^f=0$, and vice versa.  The first situation is
depicted in Figure \ref{fig_detuning}(a).  The impurity coupling $J$ is kept unchanged in this
quench protocol. 

For the ground state, tuning of the on-site impurity potential changes the structure of the impurity
cloud at small distances (within the screening cloud, $x<\xi$) and eventually destroys the
logarithmic dependence $\propto {\rm ln}~ x$ of the correlation function $\xpct{d^{\dg}c_i}$ in this
spatial region \cite{ghosh}.  However, the effect is significant only at large $\ve_d$ \cite{ghosh}.
Here, we consider moderate values of $\ve_d$, of the order of the bath bandwidth.

Figures \ref{fig_detuning}(b,c) show quenches of the detuning to a final zero value, $\ve_d^f=0$.
The relaxation of correlators to the final ground state value is exponential, with the relaxation
rate set by the screening cloud scale, $\Gamma_f= J_f^2/2$, as before.  The data shows the usual
half-bandwidth frequency oscillations. In Figure \ref{fig_detuning}(c) the oscillation periods
appear to be differentfor different $J$, but this is only due to the $J$-dependent scaling of the
time axis.

In Figures \ref{fig_detuning}(d,e) we show the dynamics after quenches to $\mathcal{O}(1)$ values of
$\ve_d^f$.  The decay to the final values occur exponentially, with the same time scale $\Gamma_f$,
consistent with the fact that moderate $\ve_d^f$ values do not destroy the screening cloud
\cite{ghosh}.  However, there are now strong oscillations with freuquency $\ve_d^f$.  To highlight
the frequencies, we use axes without dividing by the screeing scale.  The correlators now show
oscillations with two frequencies --- one equal to half the bandwidth and one equal to $\ve_d^f$.
In Figure \ref{fig_detuning}(d) the new frequency $\ve_d^f=1$ (prominient oscillations at smaller
times) is half that of the smaller oscillations visible at later times.  In \ref{fig_detuning}(e)
the oscillation frequency visibly varies with $\ve_d^f$.

At extremely large final detuning values, $\ve_d^f\gg1$ (not shown), the relaxation appears to be
power-law-like rather than exponential, for accessible system sizes.  Since this value is larger
than the bandwidth, this is beyond the regime of usual impurity phsyics, and is consistent with the
absence of impurity screening.

\section{Global quenches with local effects  \label{sec_global}}

\begin{figure}
\begin{center}
\includegraphics[width=0.99\linewidth]{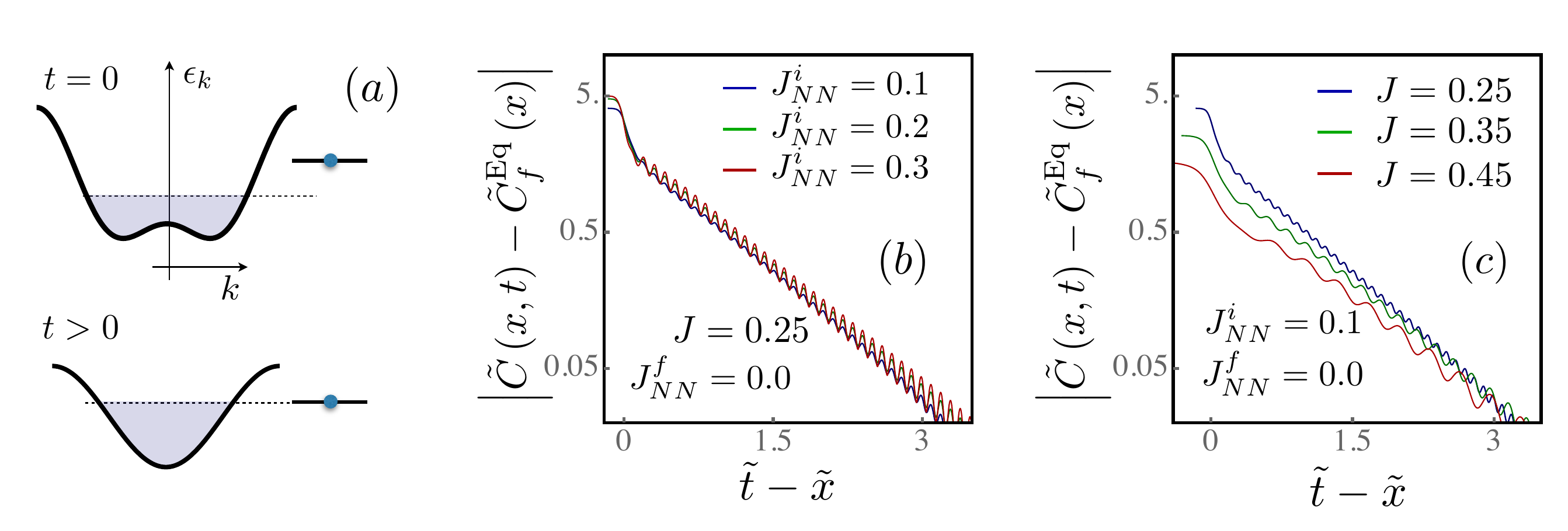}
\end{center}
\caption{  
(a) Schematic showing a quench from finite $J_{NN}$ to $J_{NN}=0$. 
(b,c) Relaxation of impurity-bath correlators, for various quenches ending at  $J^{f}_{NN}=0$.
Panel (b) uses various initial $J_{NN}$ values and panel (c) uses various $J$ values. The behavior
is remarkably similar to the case of detuning quenches. 
 }
\label{fig_global}
\end{figure}

In this section we consider a \emph{global} quench which behaves effectively like a local quench, in
the sense that the correlators relax to the ground state impurity screening cloud corresponding to
the final Hamiltonian.  We consider a quench that changes the dispersion of the bath, through a
next-nearest neighbor hopping with opposite sign compared to the nearest-neighbor hopping.  The
Hamiltonian is now
\begin{eqnarray}  \label{eq_H_lattice_nn}
H =  -\sum_{i=-L+1}^{L-1}(c_i^{\dagger}c_{i+1}+ c_{i+1}^{\dagger}c_{i}) 
~+~ J_{NN} \sum_{i=-L+1}^{L-1}(c_i^{\dagger}c_{i+2}+ c_{i+2}^{\dagger}c_{i}) 
\nonumber \\ 
~+~ \ve_d d^{\dg}d  ~-~ J\left(c_0^{\dagger}d ~+~ d^{\dagger}c_0\right).
\end{eqnarray}
This corresponds to a dispersion relation $\varepsilon_k= - 2J \cos(k) + 2 J_{NN} \cos(2k)$, as
shown in the top cartoon in Figure \ref{fig_global}(a).   

The equilibrium (zero temperature) effects of such a modification to the bath dispersion was
explored in some detail in Ref.\ \cite{ghosh}.  At any filling fraction, there are three regimes of
behavior depending on the value of $J_{NN}$.  For half-filling these behaviors are as follows: For
$J_{NN}>1$, the bath has two pairs of Fermi points rather than a single pair; the impurity screening
cloud has the same form except that the density of states now has contributions from four Fermi
points.  The $J_{NN}=1$ case is special as there is single Fermi `touching' point instead of a pair
of Fermi crossing points.  In this case the correlators show a non-Fermi-liquid character for
distances exceeding the screening length.  Finally, the $0<J_{NN}<1$ regime is very similar to the
case of the un-modified bath, with two Fermi points. 

In the absence of the impurity, the single-particle Hamiltonian with and without the $J_{NN}$ term
has the same eigenstates for each momentum, only the eigen-energies are affected.  For any
$J_{NN}\in[0,1]$, the same single-particle momenta $k\in[-\frac{\pi}{2},\frac{\pi}{2}]$ constitute
the ground-state Fermi sea at half filling, so that an abrupt change of $J_{NN}$ in the absence of
an impurity would leave the system stationary and not have any non-equilibrium effects.  Such a
quench has an effect only due to the presence of an impurity, so that it is effectively a local
quench. 
For present purposes we consider half-filling and $0\leq J_{NN}< 1$.  

Figure \ref{fig_global} shows data for quenches from nonzero $J_{NN}$ to $J_{NN}=0$, with the
detuning parameter set to zero: $\ve_d=0$.  The initial equilibrium state has two Fermi points as in
the case of an un-modified bath; however the chemical potential is no longer at zero for
half-filling, so that the detuning compared to the Fermi surface is initially nonzero.  Such a
quench is thus effectively a detuning quench of the impurity.  In fact, the results for the time
evolution of impurity-bath correlators shown in Figure \ref{fig_global}(b,c) are remarkably similar
to those in Figures \ref{fig_detuning}(b,c), where the detuning was quenched from nonzero values to
zero.  The change in the dispersion modifies $v_F$ and $\rho(\ve_F)$; this also has an effect on the
dynamics of impurity-bath correlators, but the dominant effect is clearly the change of detuning
between Fermi level and impurity energy.  

We have also performed quenches from $J_{NN}=0$ to $J_{NN}>0$ (not shown).  Repeating the argument
above, these are similar to detuning quenches from zero to nonzero detuning.  Accordingly, the time
evolution of $C(x,t)= \xpct{d^{\dagger}c_x}$ is very similar to the detuning quench results of
Figures \ref{fig_detuning}(d,e).

\newpage

\section{Discussion}

We have presented a study of the dynamics of the impurity screening cloud in the resonant level
model following local (or effectively local) quenches.  We have focused mostly on the
time-dependence of the equal-time impurity-bath correlators $C(x,t) = \xpct{d^{\dg}(t) c_x(t)}$,
which characterizes the screening cloud at equilibrium \cite{ghosh}.

Due to the finite speed of propagation in the bath, the effect of the quench is felt at point $x$
only at times $t>x/v_F$.  In terms of the correlator $C(x,t)$, this means that for $t<x/v_F$ the
dynamics is due only to the evolution of the $d^{\dg}$ operator.  Thus, $C(x,t)$ has relatively
simple evolution for $t<x/v_F$.  After the information wavefront has passed point $x$, the
correlator relaxes to the ground state of the final Hamiltonian, provided the system is large
enough.  We have outlined how the behavior is modified for finite-size baths, and how the
infinite-system quench properties can be extracted from time evolution data on finite-size systems.

The relaxation to the final ground state value is found to happen \emph{exponentially}, with the
rate given by the energy scale $\Gamma_f$ associated with impurity screening in the final
Hamiltonian.  This is the scale of broadening of the impurity spectral function \cite{ghosh},
analogous to the Kondo temperature scale in the single-impurity Kondo model.  We have highlighted
the relaxation with this scale after the wavefront passes the point at time $x/v_F$, by plotting
$\left|C(x,t)-C_f^{Eq}(x)\right|$ against the rescaled and shifted time variable
$\Gamma_f(t-\frac{x}{v_F})= \tilde{t}-\tilde{x}$ for a variety of quenches, with $\Gamma_f$
determined by the final impurity coupling $J_f$.  The relaxation curves plotted in this way look
broadly similar for a range of quenches.  For the case of a quench starting from $J_i=0$, we derived
and presented an explicit expression for $C(x,t)$, which analytically expresses these physical
effects, e.g., through explicit appearance of $\theta(\tilde{t}-\tilde{x})$ factors signaling the
information wavefront.   

We have also described the effect of a detuning quench on the equal-time correlators.  For moderate
detunings, the exponential relaxation persists, and if there is a final nonzero detuning with
respect to the chemical potential, this provides an additional oscillation frequency.  A quench of
the bath involving nearest-neighbor hoppings reduces effectively to a local quech as long as the
number of Fermi points is preserved.

The evolution of the entanglement entropy of a block centered around the impurity shows a clear
light-cone effect (no dynamics before $t\sim{x}/v_F$).  For $t>x/v_F$, we find two relaxational
regimes.  The first regime shows exponential relaxation with the rate $\Gamma_f$.  The longer-time
regime is consistent with a power-law relaxation with logarithmic corrections, as in the case of
Ref.\ \cite{Eisler_Peschel}.

The dynamics of impurity screening clouds has been the subject of a few recent studies
\cite{Nuss2014, Lechtenberg2014, Medvedyeva}, complementary to ours.  We briefly outline relevant
parts of these results below and compare with our results.

In Ref.\ \cite{Medvedyeva}, the authors have studied the Kondo model at the so-called Toulouse point
\cite{toulouse}.  At the Toulouse point, the Kondo model reduces to the non-interacting resonant
level model, and hence various analytical results can be derived in the case when one starts from
zero coupling.  The authors focus on spin-spin correlators of the Kondo model, which translate to
different fermionic correlators from the ones occurring naturally in this work.  Focusing on
unequal-time correlators, Ref.\ \cite{Medvedyeva} distinguishes correlators which show a strict light
cone from those which do not. 

In Ref.\ \cite{Lechtenberg2014}, quenches in the isotropic Kondo model are studied using
time-dependent NRG and perturbation theory, for both ferromagnetic and antiferromagnetic Kondo
couplings.  The time evolution of equal-time impurity-bath spin-spin correlators are studied.  As in
our case, the correlators considered do not have a strict light cone.

Ref.\ \cite{Nuss2014} addresses several issues similar to the ones addressed in the present work.
In this work, the dynamics of the Anderson model is studied with a focus on the Kondo regime.  Two
types of equal-time impurity-bath correlators are studied: spin-spin and density-density
correlators.  The behavior for $t<x/v_F$ and $t>x/v_F$ are characterized in some detail, and the
exponential decay for $t>x/v_F$ is discussed.  The decay rate appears to be more complicated than in
our case and weakly dependent on position; it is not completely clear whether this is due to system
size issues.

The present study has focused on the noninteracting RLM and the dynamics of correlators naturally
suitable for describing the RLM screening cloud.  The absence of interactions has allowed us to
numerically treat relatively large systems, making possible an unambiguous identification of the
time scale of relaxation.  This aspect is more difficult with sizes accessible for interacting
systems \cite{Nuss2014}.  In addition, our analytical result for the $J_i=0$ case explicitly shows
${\sim}e^{-\Gamma_ft}$ relaxation of the equal-time impurity-bath correlators.  Based on the present
results, it is natural to conjecture that the relaxation rate is given by the characteristic
equilibrium impurity energy scale for any impurity model characterized by an emergent energy scale.
For example, the Kondo temperature scale should provide the relaxation rate for the
antiferromagnetic Kondo case of Ref.\ \cite{Lechtenberg2014}, but the details of relaxation are not
yet clarified in the literature.

The previous studies concentrate on initially decoupled impurities, i.e., cases where there is no
nontrivial screening cloud in the initial state.  In section \ref{subsec_Ji_finite} we have
presented numerical results for finite initial coupling $J_i$.  Even though the relaxation is
governed solely by the final impurity energy scale, the structure of the initial screening cloud
induces strong oscillations superposed on the relaxation.

Local quenches are being currently studied from a variety of perspectives.  The present work,
together with the related literature discussed above, addresses the role of the emergent energy
scale in the dynamics, in cases where the local feature possesses such a scale.  Several aspects
remain poorly understood and clearly deserve further investigation and explanation, such as the
behavior of the block entanglement entropy and entanglement spectrum, the role of finite bandwidths,
etc.

\bigskip

\ack

The authors thank I.~Affleck, E.~Boulat, M.~Ganahl, F.~Heidrich-Meisner, S.~Kehrein, M.~Medvedyeva,
M.~Nuss, Y.~Shchadilova, and R.~Vasseur for useful discussions.

\bigskip

\section{Details of the derivation}

Eq.\ (\ref{dcanalytic}), reported in the main text, gives the time evolution of impurity-bath
correlators in the infinite-bandwidth and infinite-bath-size limit.  In this appendix we briefly
outline the derivation of this expression.

The evolution of the system after the quench at time $t=0$ is governed by the final Hamiltonian
given in Eq.(\ref{H_RLM}) which in its eigen-operator basis $a_\varepsilon$ can be written in the
form $H=\sum_\varepsilon \varepsilon  \, a^\dagger_\varepsilon a_\varepsilon$.  
The new set of operators $a_{\ve}$, $a_{\ve}^{\dagger}$ can be related with the original $c_k$ and
$d$ by imposing the commutation relations $\left[H, a_\varepsilon \right] = - \ve a_\varepsilon $.
This yields 
\begin{equation} 
a_{\ve}=A_{\ve d } d + \sum_k A_{\ve k} c_k, \quad  
a_{\ve}^{\dg}=A_{\ve d }^{*} d^{\dg} + \sum_k A_{\ve k}^{*} c_k^{\dg}
\end{equation}
with 
\begin{eqnarray}
A_{\ve d} &=& A_{\ve d}^{*}= \sqrt{\frac{2v_F}{2L+1}\frac{\Gamma}{\Gamma^2+\ve^2}},\\
A_{\ve k} &=& A_{\ve k}^{*} = \frac{J'}{\sqrt{2L+1}}\frac{1}{\ve-\ve_k}\sqrt{\frac{2v_F}{2L+1}\frac{\Gamma}{\Gamma^2+\ve^2}}
\end{eqnarray}
Note that the unitarity of the transformation also implies that
\begin{equation} 
c_k=\sum_{\ve} A_{\ve k} a_{\ve}; \quad d= \sum_{\ve} A_{\ve d} a_{\ve}. 
\end{equation}

The time evolution of the operators $d^{\dg}(t)$ and $c_r(t)$ can be most easily computed in the
basis where $H$ is diagonal.  This leads to the expressions:
\begin{eqnarray}
d^{\dg}(t) &=& \sum_{\ve} |A_{\ve d}|^2 e^{i\ve t} d^{\dg}(0)+\sum_{\ve,k}A_{\ve d}A_{\ve k}e^{i\ve
  t} c_k^{\dg}(0),  \label{dt} \\ 
c_x(t) &=& \frac{1}{\sqrt{2L+1}} \Bigg[ \sum_{k,\ve} A_{\ve k} A_{\ve d}  e^{ikx} e^{-i\ve t} d(0) \nonumber
\\ &+& \sum_{k,k',\ve} A_{\ve k} A_{\ve k'}  e^{ikx} e^{-i\ve t} c_{k'}(0 
)\Bigg],  \label{crt}
\end{eqnarray}
relating the time evolved operators to the operators at $t=0$.  In order to obtain a useful
expression for the correlator $\xpct{d^{\dg}(t) c_x(t)}$, we now introduce two approximations.
The first is the thermodynamic limit where an infinite chain is considered. 
The second is the wide-band approximation where we assume that the energy scales of interest are much smaller than the bandwidth of the chain. 
This amounts to considering that the density of states of the bath is constant.  With these two
approximations in place, the summations over momentum can be done assuming a linear dispersion
relation around the two Fermi points: $\varepsilon_k = v_F (k-k_F) $ and $\varepsilon_k = - v_F
(k+k_F)$.  Thus for any function $f(k)$ of momentum we have: 
\begin{eqnarray}
\frac{1}{2L+1} \sum_k f(k) \to \int \frac{d\varepsilon_k} {2\pi v_F}  \left[ f(  k_F + v_F^{-1} \varepsilon_k  ) + f(  - k_F - v_F^{-1} \varepsilon_k  ) \right]
\end{eqnarray}
However, some of the coefficients of the unitary transformation have denominators of the form $\frac{1}{\ve-\ve_k}$ which have to be treated with care when $\ve \simeq \ve_k$. 
By analysing the discrete version of the eigenvalue equation of the $H$:  
\begin{eqnarray}
\varepsilon = \frac{J^2}{2L+1}\sum_k \frac{1}{\ve-\ve_k}
\end{eqnarray}
one arrives at the prescription (see also Ref.\cite{Medvedyeva} and references therein)
\begin{eqnarray}
\frac{1}{\ve-\ve_k} \to P\left(  \frac{1}{\ve-\ve_k} \right) + \pi \frac{\ve}{\Gamma} \delta \left(  \ve-\ve_k \right). 
\end{eqnarray}
where $P$ stands for the principle value. 
Within the considered approximations, the summations in Eqs.(\ref{dt}, \ref{crt}) can thus be
replaced by integrals and partially done by contour deformation in the complex plane. An example of
such a procedure for a typical term arising in the computation of $c_x(t)$ is: 
\begin{eqnarray}
\frac{1}{\sqrt L}\sum_{\varepsilon, k}  e^{i(kx- \varepsilon t )}  A_{\varepsilon k} A_{\varepsilon d} & \simeq &  \frac{J}{2\pi^2} \int d\ve \frac{e^{-i\ve t}}{\ve^2+\Delta^2}  \sum_{s=\pm 1} \int \frac{d\varepsilon_k} {2\pi v_F} 
e^{is(k_F + v_F^{-1} \varepsilon_k)x}  \times \non \\ 
&& \left[{\rm P}\left(\frac{1}{\ve-\ve_k} \right)+\pi\frac{\ve}{\Delta}\delta(\ve-\ve_k) \right] \non  \\
& = & -i \frac{J}{v_F} e^{i k_F x}  \theta\left( \tilde t - \tilde x  \right) e^{- (\tilde t - \tilde x) }  
\end{eqnarray}
where we assumed $x>0$ for simplicity. 
The pole structure of the expression leads to the theta-function term  $\theta(\tilde t - \tilde x)$ in Eq.(\ref{dcanalytic})  that signals the passage of the wave front through the rescaled position $\tilde x$.


\newpage


\section*{References}


\begin{thebibliography}{99}





\bibitem{Medvedyeva} Medvedyeva M, Hoffmann A, and Kehrein S, 2013 Phys.\ Rev.\ B \textbf{88}, 094306 

\bibitem{Lechtenberg2014} Lechtenberg B and Anders F B, 2014 Phys.\ Rev.\ B {\bf 90} 045117 

\bibitem{Nuss2014} Nuss M,  Ganahl M, Arrigoni E, von der Linden W, and Evertz H G, 2014  Phys.\ Rev.\ B \textbf{91}, 085127



\bibitem{KleineAnders_PRB2014} Kleine C, Mu{\ss}hoff J, and Anders F B, 2014 Phys.\ Rev.\ B
  \textbf{90}, 235145
%



\bibitem{VasseurHaasSaleur_PRL2013}  Vasseur R, Trinh K, Haas S, and Saleur H, 2013  Phys.\ Rev.\ Lett.\ \textbf{110}, 240601 
%

\bibitem{VasseurMoore_PRL2014}     Vasseur R and Moore J E, 2014 Phys.\ Rev.\ Lett.\ \textbf{112}, 146804
%


\bibitem{KennesMedenVasseur_PRB2014}  Kennes D M, Meden V, and Vasseur R, 2014 Phys.\ Rev.\ B \textbf{90}, 115101
%

\bibitem{SchillerAnders_PRB2014}  Vinkler-Aviv Y, Schiller A, and Anders F B, 2014
Phys.\ Rev.\ B  \textbf{90}, 155110 
%

\bibitem{Latta} Latta C, Haupt F, Hanl M, Weichselbaum A, Claassen M, Wuester W, Fallahi P, Faelt S,
  Glazman L, von Delft J, T\"ureci H E, and Imamoglu A, 2011 Nature {\bf 474} 627
%


\bibitem{Schiro_PRB2012} Schir\'o M, 2012  Phys.\ Rev.\ B \textbf{86} 161101(R) 
%



\bibitem{vonDelft_theory_PRB2012}  M\"under W, Wechselbaum A, Goldstein M, Gefen Y, and von Delft J,
  2012 Phys.\ Rev.\ B \textbf{85}, 235104  
%


\bibitem{SchiroMitra_PRL2014} Schir{\'o} M and Mitra A,  2014  Phys.\ Rev.\ Lett.\ \textbf{112}, 246401 
%


\bibitem{SchillerAndrei_PRB2014}
Vinkler-Aviv Y, Schiller A, and  Andrei N, 2014  Phys.\ Rev.\ B \textbf{89}, 024307


\bibitem{Costi_PRB14} Nghiem H T M and Costi T A,  2014 Phys.\ Rev.\ B \textbf{89}, 075118
%
%
\\ Nghiem H T M and Costi T A,  2014 Phys.\ Rev.\ B \textbf{90}, 035129
%


\bibitem{RatianiMitra_PRB2010}  Ratiani Z and Mitra A,  2010  Phys.\ Rev.\ B \textbf{81}, 125110 
%


\bibitem{HeylVojta_PRL2014}
Heyl M and Vojta M, 2014 Phys.\ Rev.\ Lett.\ \textbf{113}, 180601 


\bibitem{HeylKehrein_JPCM2010}
Heyl M and Kehrein S, 2010  J.~Phys.: Condens.\ Matter \textbf{22}, 345604 
%


\bibitem{LobaskinKehrein_PRB2005} Lobaskin D and Kehrein S, 2005 Phys.\ Rev.\ B \textbf{71}, 193303 




\bibitem{kondo} Kondo J, 1964 Prog. Theo. Phys. {\textbf 32}, 37 

\bibitem{hewson} Hewson A C, 1993 \textit{The Kondo Problem to Heavy Fermions} (Cambridge University  Press, New York)

\bibitem{Anderson_model_PRB61}  Anderson P W, 1961 Phys.\ Rev.\ \textbf{124} 41 

\bibitem{Anderson_model_scale} Haldane F D M, 1978  J.~Phys.\ C \textbf{11}, 5015 
%
\\ Zitko R, Bonca J, Ramsak A, and Rejec T, 2006 Phys.\ Rev.\ B \textbf{73}, 153307 


\bibitem{IRLM_original} Schlottmann P, 1980 Phys.\ Rev.\ B \textbf{22}, 613 
%
\\  Schlottmann P, 1982 Phys.\ Rev.\ B \textbf{25}, 4815 
%
\\  Filyov V M, Tzvelik A M, and Wiegmann P B, 1980 Phys.\ Lett.\ A \textbf{81}, 175


\bibitem{IRLM_recent} Boulat E and Saleur H, 2008 Phys.\ Rev.\ B \textbf{77}, 033409
%
\\ Boulat E, Saleur H, and Schmitteckert P, 2008 Phys.\ Rev.\ Lett.\ {\bf 101},
  140601



\bibitem{AffleckSimon_PRL01} Affleck I and Simon P, 2001 Phys. Rev. Lett. \textbf{86}, 2854 

\bibitem{SimonAffleck_PRB01}   Simon P and Affleck I, 2001 Phys.\ Rev.\ B {\bf 64}, 085308 

\bibitem{SimonAffleck_PRL02}   Simon P and Affleck I, 2002 Phys.\ Rev.\ Lett.\ {\bf 89}, 206602

\bibitem{SimonAffleck_PRB03}   Simon P and Affleck I, 2003 Phys.\ Rev.\ B {\bf 68}, 115304 

\bibitem{HandKrohaMonien_PRL06} Hand T, Kroha J, and Monien H, 2006 Phys.\ Rev.\ Lett.\ {\bf 97}, 136604 



\bibitem{BusserAndaDagotto_PRB10}  B\"usser C A, Martins G B, Costa~Ribeiro L, Vernek E,
  Anda E V, and Dagotto E, 2010 Phys. Rev. B \textbf{81}, 045111

\bibitem{Simonin_arXiv07}  Simonin J, arXiv:0708.3604.

\bibitem{Goth_Assaad} Goth F, Luitz D J, and Assaad F F, 2013 Phys. Rev. B \textbf{88}, 075110


\bibitem{Holzner-etal_PRB09} Holzner A, McCulloch I P, Schollwöck U, von Delft J, and
  Heidrich-Meisner F,   2009 Phys. Rev. B \textbf{80}, 205114

\bibitem{Borda} Borda L, 2007 Phys. Rev. B {\textbf 75}, 041307(R) 

\bibitem{BordaGarstKroha_PRB09} Borda L, Garst M, and Kroha J, 2009 Phys.\ Rev.\ B \textbf{79}, 100408(R)

\bibitem{Saleur}  Affleck I, Borda L, and  Saleur H, 2008 Phys. Rev. B  {\bf 77}, 180404(R)

\bibitem{MitchellBulla_PRB11} Mitchell A K, Becker M, and Bulla R, 2011  Phys.\ Rev.\ B \textbf{84}, 115120


\bibitem{SougatoBose_kondocloud} A.~Bayat, P.~Sodano, and S.~Bose, Phys.\ Rev.\ B {\bf 81}, 064429 (2010). 


\bibitem{Laflorencie_JSM1} S{\o}rensen E S, Chang M S, Laflorencie N, and Affleck I, 2007
  J.~Stat.\ Mech.\  P08003.

\bibitem{Laflorencie_JSM2} S{\o}rensen E S, Chang M S, Laflorencie N, and Affleck I, 2007
  J.~Stat.\ Mech.\  L01001.


\bibitem{BarzykinAffleck_PRB98} V.~Barzykin and I.~Affleck, Phys.\ Rev.\ B {\textbf 57}, 432 (1998).

\bibitem{ishii} H. Ishii, J. Low Temp. Phys. {\textbf 32}, 457 (1978).

\bibitem{SorensenAffleck96}  S{\o}rensen E S and Affleck I, 1996 Phys.\ Rev.\ B {\textbf 53}, 9153 



\bibitem{ghosh} Ghosh S,  Ribeiro P and Haque M, 2014 J.~Stat.\ Mech.\ P04011 


\bibitem{SaleurVasseur_PRB13} Saleur H, Schmitteckert P, and Vasseur R, 2013 Phys.\ Rev.\ B
  \textbf{88}, 085413.





 
\bibitem{LiebRobinson_1972}
Lieb E H and Robinson D, 1972 \emph{Commun. Math. Phys.} \textbf{28} 251

\bibitem{Nachtergaele_JSP2006}
Nachtergaele B, Ogata Y, and Sims R, 2006 J. Stat. Phys. \textbf{124} 1
%

\bibitem{CalabreseCardy_PRL2006}
Calabrese P and Cardy J, 2006 Phys. Rev. Lett. \textbf{96}, 136801 
%

\bibitem{BravyiHastingsVerstraete_PRL2006}
Bravyi S, Hastings M B, and Verstraete F, 2006 Phys.\ Rev.\ Lett.\ \textbf{97}, 050401 


\bibitem{NachtergaeleSims_CMP2006} Nachtergaele B and Sims R, 2006 Commun.\ Math.\ Phys.\
  \textbf{265}, 119

\bibitem{SchuchEisert_PRA2011} 
Schuch N, Harrison S K, Osborne T J, and Eisert J, 2011 Phys.\   Rev.\  A \textbf{84}, 032309 


\bibitem{EsslerLauchli_PRL2014}
Bonnes L, Essler F H L, and L\"auchli A M, 2014 
Phys.\ Rev.\ Lett.\ \textbf{113}, 187203
%


\bibitem{ChenauBlochKollath_Nature2012}
Cheneau M, Barmettler P, Poletti D, Endres M, Schau{\ss} P, Fukuhura T, Gross C, Bloch I, Kollath~C,
and Kuhr S, 2012 Nature (London) \textbf{481}, 484


\bibitem{StephanDubail_JSM2011}
St{\'e}phan J-M and Dubail J,  2011 J.~Stat.\ Mech.\ P08019


\bibitem{DivakaranIgloiRieger_local_JSM2011}  Divakaran U, Igl{\'o}i F, Rieger H, 2011 J.~Stat.\ Mech.\ P10027
%


\bibitem{LewensteinTagliacozzo_JSM2014}
Zamora A, Rodr{\'i}guez-Laguna J, Lewenstein M, and Tagliacozzo L, 2014 J.~Stat.\ Mech.\ P09035
%


\bibitem{Manmana_etal_PRB2009}
Manmana S R, Wessel S, Noack R M, and Muramatsu A, 2009
Phys.\ Rev.\ B  \textbf{79}, 155104
%


\bibitem{Cazalilla_PRL2006}  Cazalilla M A, 2006  Phys.\ Rev.\ Lett.\ \textbf{97}, 156403 





\bibitem{Eisler_Peschel}  Eisler V and Peschel I, 2007 J.~Stat.\ Mech.: Theory Exp.  P06005.

\bibitem{Peschel_JPA2003}  Peschel I, 2003 J.~Phys.~A: Math.\ Gen.\ \textbf{36}, L205







\bibitem{toulouse} Toulouse G, Phys.\ Rev.\ B {\textbf 2}, 270 (1970). 









\end{thebibliography}
\end{document}